\documentclass[a4paper]{jpconf}

\usepackage{graphicx}

\usepackage{citesort}

\begin{document}
\title{sPHENIX: The next generation heavy ion detector at RHIC}

\author{Sarah Campbell for the sPHENIX Collaboration}

\address{Columbia University, New York, NY USA}

\ead{sc3877@columbia.edu}

\begin{abstract}
sPHENIX is a new collaboration and future detector project at Brookhaven National Laboratory's Relativistic Heavy Ion Collider (RHIC). It seeks to answer fundamental questions on the nature of the quark gluon plasma (QGP), including its temperature dependence and coupling strength, by using a suite of precision jet and upsilon measurements that probe different length scales of the QGP. This will be achieved with large acceptance, $|\eta| < 1$ and $0$-$2\pi$ in $\phi$, electromagentic and hadronic calorimeters and precision tracking enabled by a $1.5$~T superconducting magnet. With the increased luminosity afforded by accelerator upgrades, sPHENIX will perform high statistics measurements extending the kinematic reach at RHIC to overlap the LHC's. This overlap with the LHC will facilitate better understanding of the role of temperature, density and parton virtuality in QGP dynamics and for jet quenching in particular. This talk will focus on key future measurements and the current state of the sPHENIX project.
\end{abstract}

\section{Introduction}
The goal of the sPHENIX program~\cite{proposal} is to probe the Quark-Gluon Plasma (QGP) created in heavy ion collisions at multiple length scales.  It does this through three avenues: studying the structure of jets as they evolve in the QGP, varying the jet parton (from the smallest gluon-jets, to light quark-, and then the larger, heavier charm and bottom quark-jets) and studying the sequential melting of the three upsilon states ($\Upsilon(1S)$, $\Upsilon(2S)$, $\Upsilon(3S)$).  Initially jets created in heavy ion collisions are highly virtual, traveling through the QGP, but as they fragment they reach a scale close to that of the QGP and QGP interactions are more probable.  As a result, the sensitivity of jet measurements to QGP effects is higher in lower energy jets and jets created at RHIC energies.  Similarly, the partonic composition of jets differs at RHIC and LHC energies, with a higher fraction of quark-jets available at lower jet energies at RHIC compared to the LHC.  Additional jet-parton studies require identifying heavy flavor jets. This will allow further study of collisional versus radiative energy loss and the dead cone effect.  Finally, measurements of the upsilon states at RHIC will provide insight on bottom quarkonia behavior at lower QGP temperatures, $T_{RHIC}$ is $77$\% lower than $T_{LHC}$, and with reduced $\Upsilon$ production from coalescence.  For each of these signals, it is necessary to obtain complementary measurements at both RHIC and the LHC to disentangle their respective contributions and sensitivities.  sPHENIX is the RHIC detector capable of complementing the LHC heavy ion program.

\section{The sPHENIX detector}
\begin{figure}[h]
\begin{minipage}[b]{2.4in}
\includegraphics[width=2.4in,trim={1.3cm 0 1.2cm 0.4cm},clip]{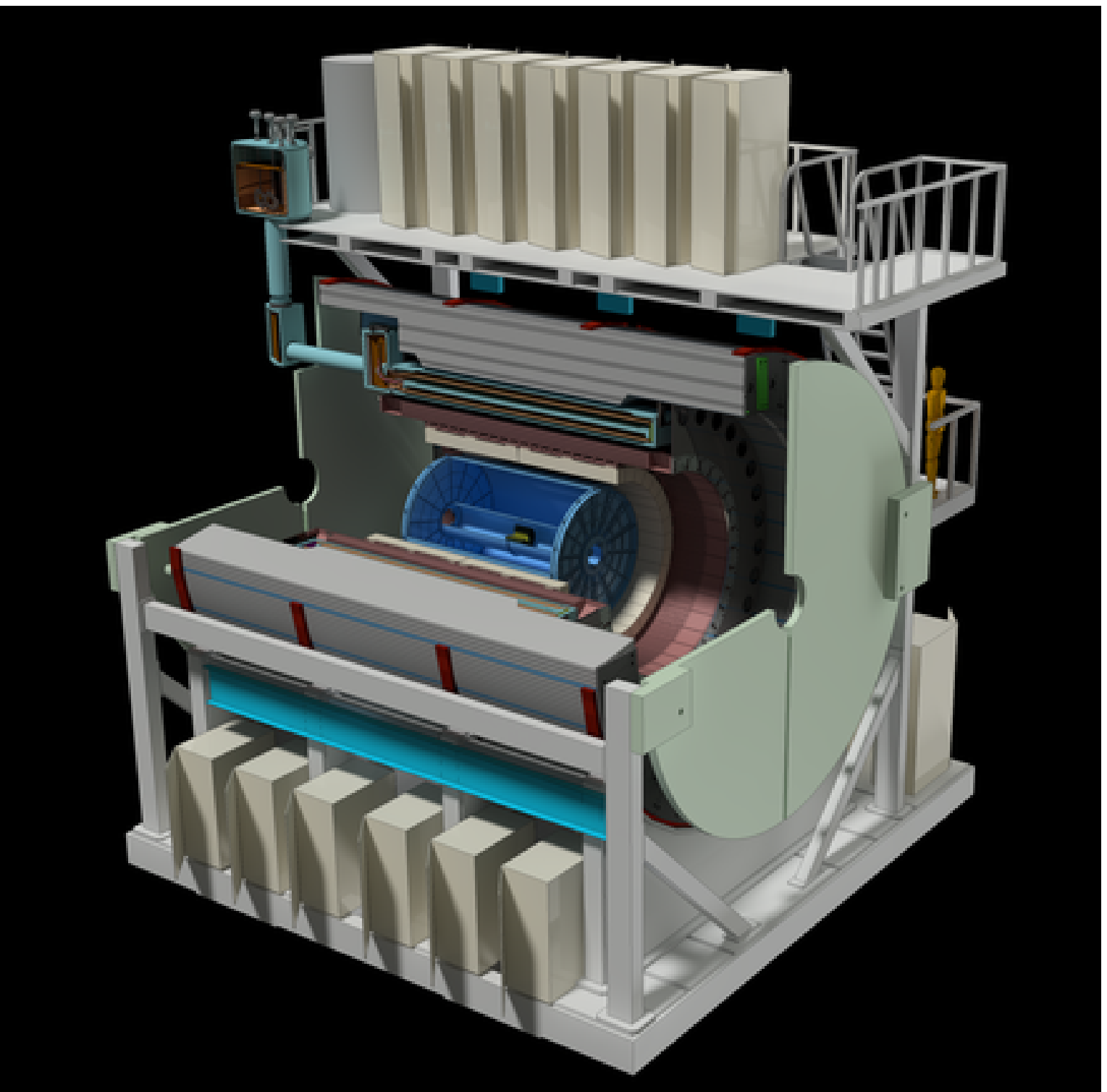}
\caption{\label{Fig:Detectors}Engineering drawing of sPHENIX with its support structure, three calorimeters, the 1.5~T magnet, and three tracking detectors.}
\end{minipage}\hspace{0.15in}
\begin{minipage}[b]{1.5in}
\includegraphics[width=1.5in,trim={7.5cm 0.6cm 8cm 1cm},clip]{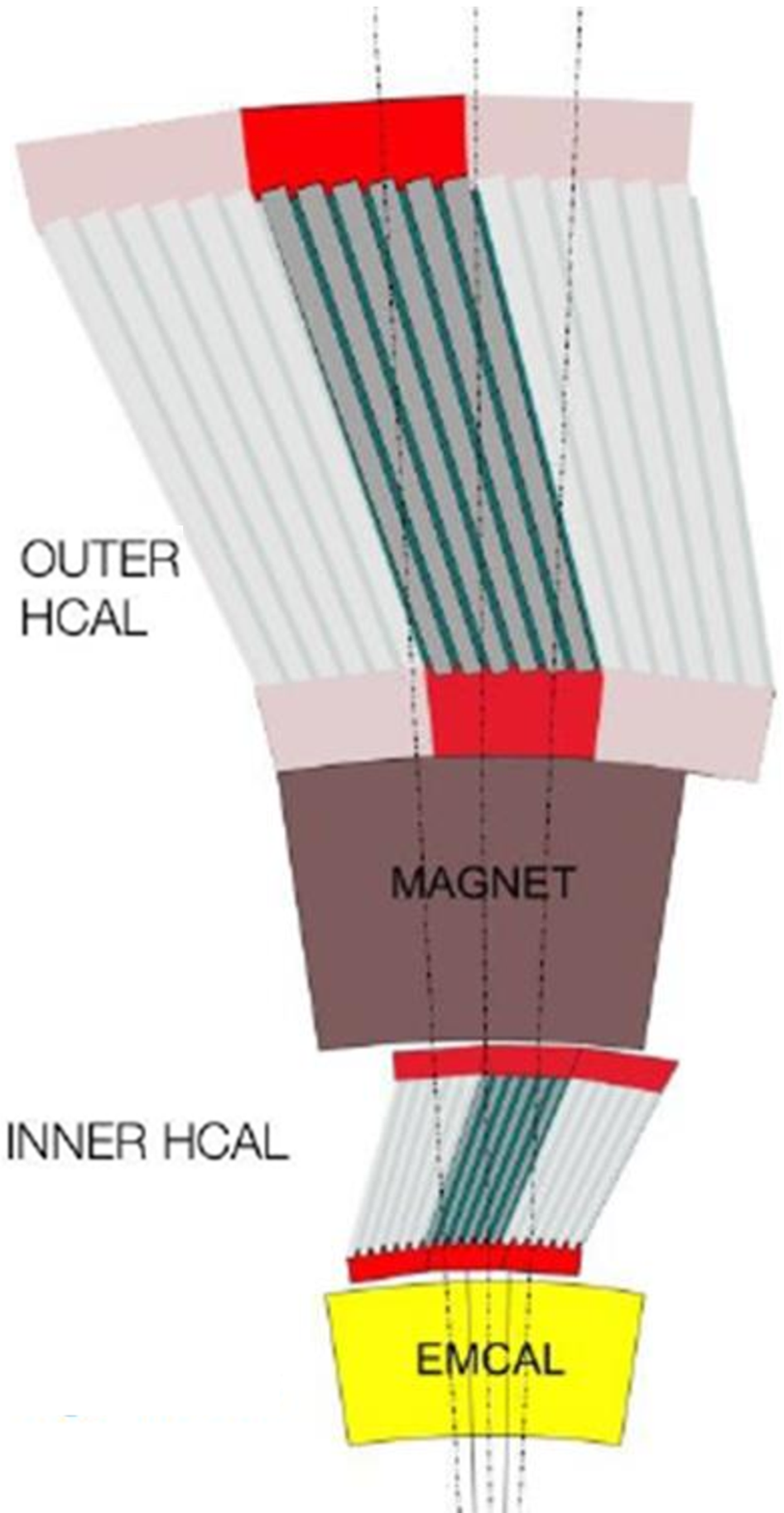}
\caption{\label{Fig:Calorimeters}Angular section of the sPHENIX hadronic and electromagnetic calorimeters shown with the 1.5~T magnet.}
\end{minipage}\hspace{0.15in}
\begin{minipage}[b]{2in}
\includegraphics[width=2in,trim={0.3cm 0 0.5cm 0},clip]{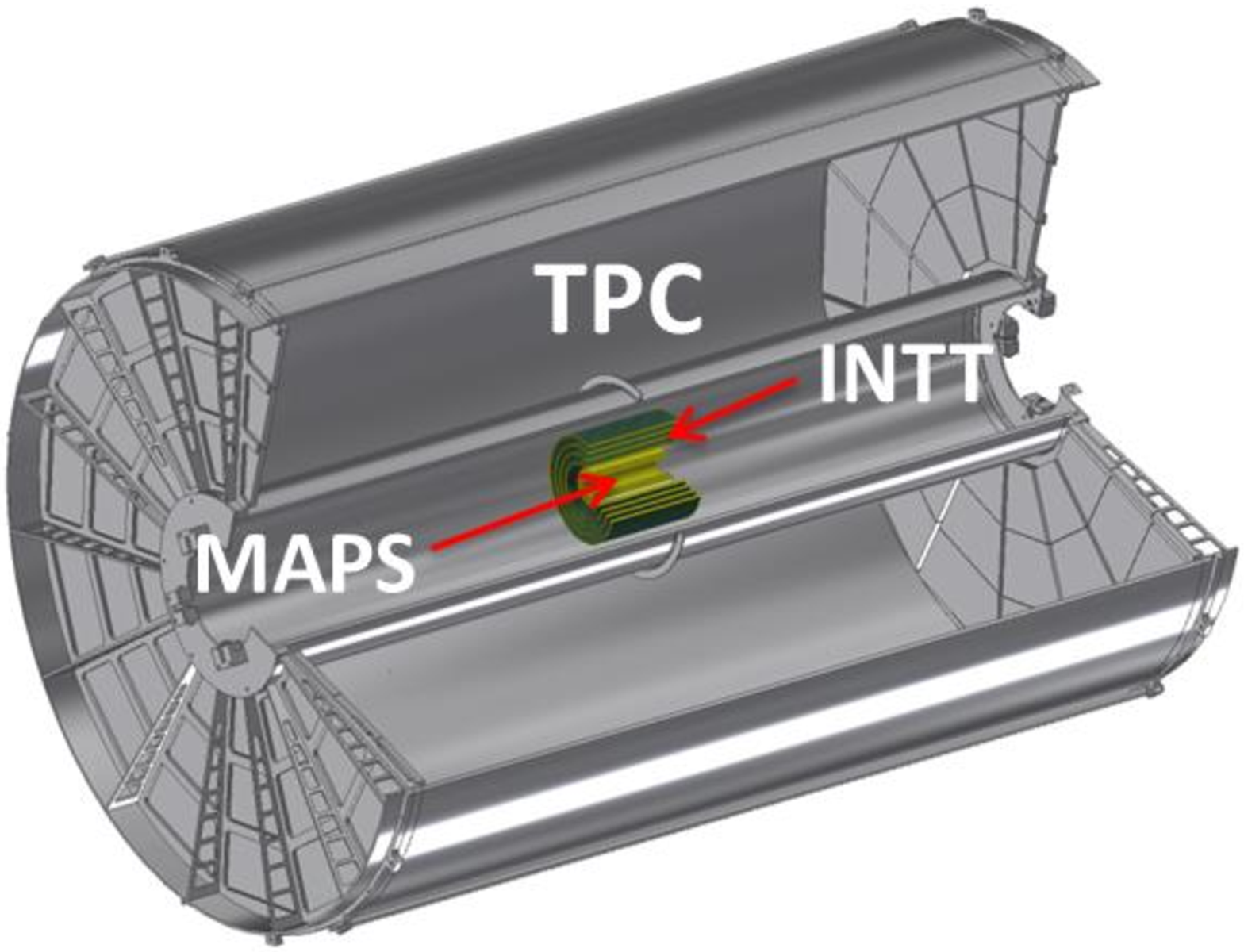}
\vspace{0.4in}
\caption{\label{Fig:Trackers}Engineering drawing of the sPHENIX tracking detectors, specifically the Time Projection Chamber (TPC), Inner Tracker (INTT), and MAPS detector.} 
\end{minipage}
\end{figure}

The sPHENIX detector design, Figure~\ref{Fig:Detectors}, is driven by the goal of measuring these rare signals.  To make the most of RHIC luminosities, the sPHENIX detector covers $2\pi$ in azimuth ($\phi$) and $\pm1$ units in rapidity ($\eta$) and reads out data at a rate of $15$~kHz.  To efficiently resolve jets and their energies, sPHENIX has both hadronic and electromagnetic calorimeters.  Efficient tracking of particles from $0.2$ to $40$~GeV/c in transverse momentum ($p_{T}$) is needed for jet fragmentation measurements.  Heavy flavor jet identification requires precision vertexing with a distance of closest approach in the $x$-$y$ plane, $DCA_{xy}$, resolution of better than $70$~$\mu$m.  To measure the three $\Upsilon$ states in the dielectron decay channel hadron rejection of better than $99\%$ and a mass resolution of $1\%$ at the $\Upsilon$ mass is required.  These precise and efficient tracking requirements are met using a $1.5$~T magnet and three tracking subsystems: a time projection chamber (TPC), a Si strip inner tracker (INTT) and a precision MAPS detector based off of ALICE's ITS upgrade.  The remainder of this section discusses the calorimeter and tracking detectors, Figures~\ref{Fig:Calorimeters} and \ref{Fig:Trackers} respectively, in more detail.
\subsection{Calorimeters}
The hadronic calorimeters (HCal) consist of tilted steel plates interleaved with polystyrene panels embedded with $1$~mm wavelength shifting fiber.  It has a segmentation of $0.1 \times 0.1$ in $\phi$ and $\eta$.  Inner and Outer HCal subsystems measure the hadronic energy deposited before and after the magnetic coil.  The HCal design has a single particle energy resolution requirement of better than $100\%/\sqrt{E}$.  The electromagnetic calorimeter (EMCal) is made up of W modules embedded with Si fibers evenly spaced in $\phi$ and $\eta$.  The EMCal $\phi$ and $\eta$ segmentation is $0.025 \times 0.025$ and it has an energy resolution requirement of better than $15\%/\sqrt{E}$.  Research and development work on the EMCal is ongoing to determine whether modules should be projective in both $\phi$ and $\eta$ or just in $\phi$.  Si-photomultipliers measure the light produced in both the HCal and EMCal systems and a common digitizer readout is planned.

In the winter of 2016, a prototype of the calorimeter systems was studied at the FermiLab Test Beam Facility.  
The measured energy distributions in the HCal are well reproduced by full GEANT simulations.  This provides added confidence in our simulated detector response.  Furthermore, preliminary analyses of the the combined HCal and EMCal single particle energy resolution and electron energy resolution in the EMCal meet the design goals, with a projected single particle energy resolution in the range $(70.6\%-95.7\%)/\sqrt{E}$ and an electron energy resolution of roughly $14.2\%/\sqrt{E}$ or $12.7\%/\sqrt{E}$ depending on the shower calibration method.

\subsection{Tracking detectors}
The outer-most tracking detector, the TPC, is located between $20$ and $78$~cm in radius and has an approximate $250$~$\mu$m effective hit resolution.  It provides the bulk of the pattern recognition and momentum resolution for the tracking of particles between $0.2$ and $40$~GeV/c in $p_{T}$.  A continuous, non-gated, TPC readout is planned to be compatible with sPHENIX's high data acquisition rate. The inner-most tracking detector, the MAPS detector, consists of three layers of Si sensors following the ALICE ITS upgrade design~\cite{ALICE}.  It contributes both precision event vertex determination, $|z_{vtx}| < 10$~cm, and identification of off-vertex decays, $DCA_{xy} < 70$~$\mu$m.  Located between the TPC and the MAPS layers, the INTT detector provides needed continuity in the tracking, redundancy in pattern recognition and DCA determination, and pile-up rejection.  It consists of $4$ layers of Si strips and will be readout out by reusing PHENIX FVTX electronics.

\section{Simulated detector performance}

The performance of the combined tracking systems is simulated by embedding pions in central HIJING events.  Current results in this ongoing effort show efficient tracking out to $40$~GeV/c in $p_{T}$ and a distance of closest approach resolution of $40$~$\mu$m at the lowest $p_{T}$ values.  The calorimetric jet reconstruction performance is characterized by the jet energy resolution, efficiency and purity.  These quantities in sPHENIX are simulated using central HIJING events with an ATLAS-influenced jet reconstruction algorithm for jet radii of $0.2$, $0.3$, and $0.4$~\cite{jets}.  The resulting jet efficiency is better than $90\%$ for jets with $p_{T}$ greater than $20$~GeV/c and the jet purity is better than $80\%$ for jets with a $p_{T}$ of greater than $25$~GeV/c.

\begin{figure}[h]
\begin{center}
\begin{minipage}{21.5pc} 
\includegraphics[width=21.5pc,trim={0.3cm 0.8cm 0 0.7cm},clip]{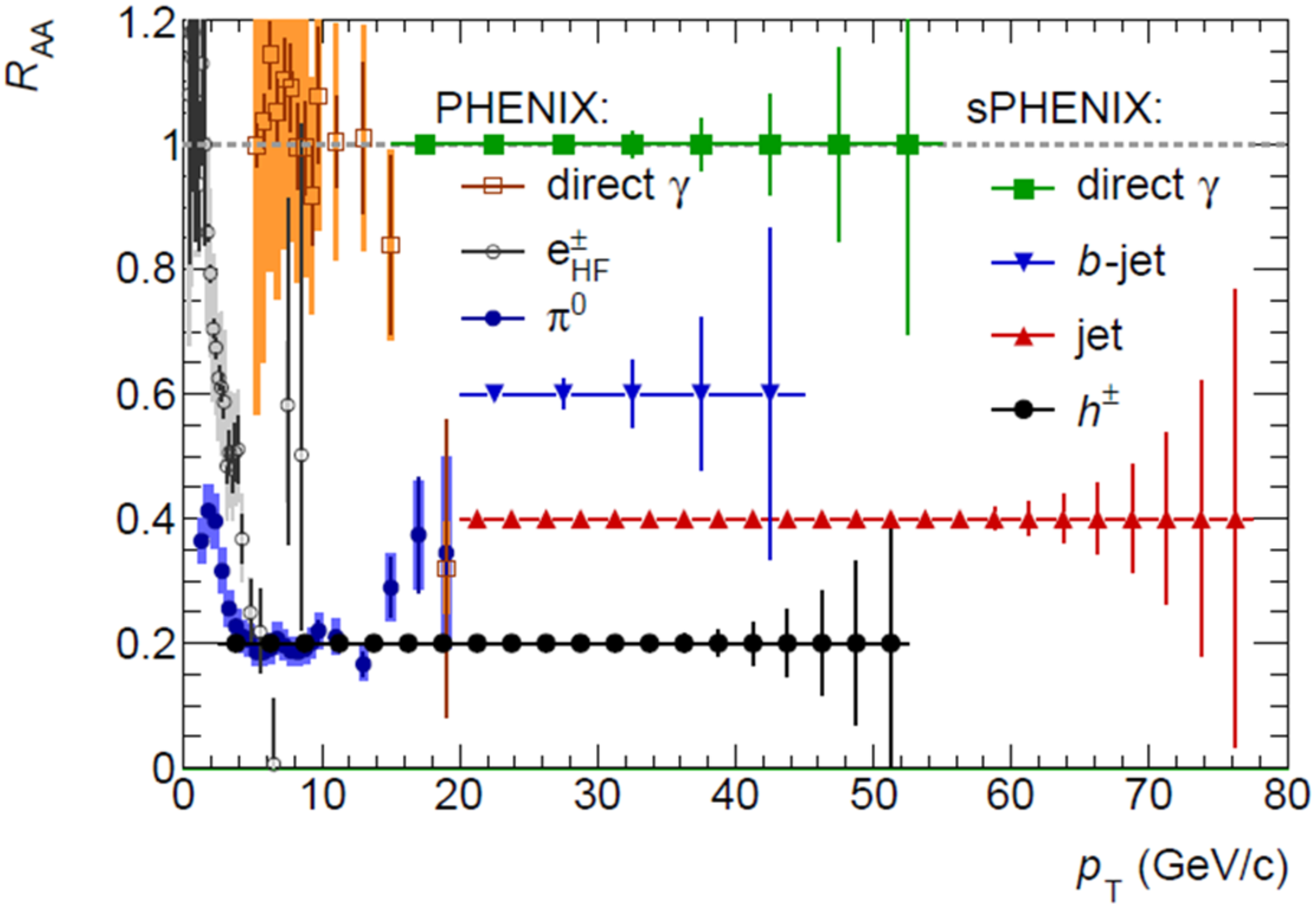}
\caption{\label{Fig:RAA}Statistical projections of the nuclear modification factor, $R_{AA}$, for various hard processes measured with sPHENIX given $22$~week of Minimum Bias $Au$+$Au$ collisions with $10$~weeks of $p$+$p$ collisions.}
\end{minipage}\hspace{1.5pc}
\begin{minipage}{13.5pc} 
\includegraphics[width=13.5pc, trim={4cm 0cm 4cm 2.1cm},clip]{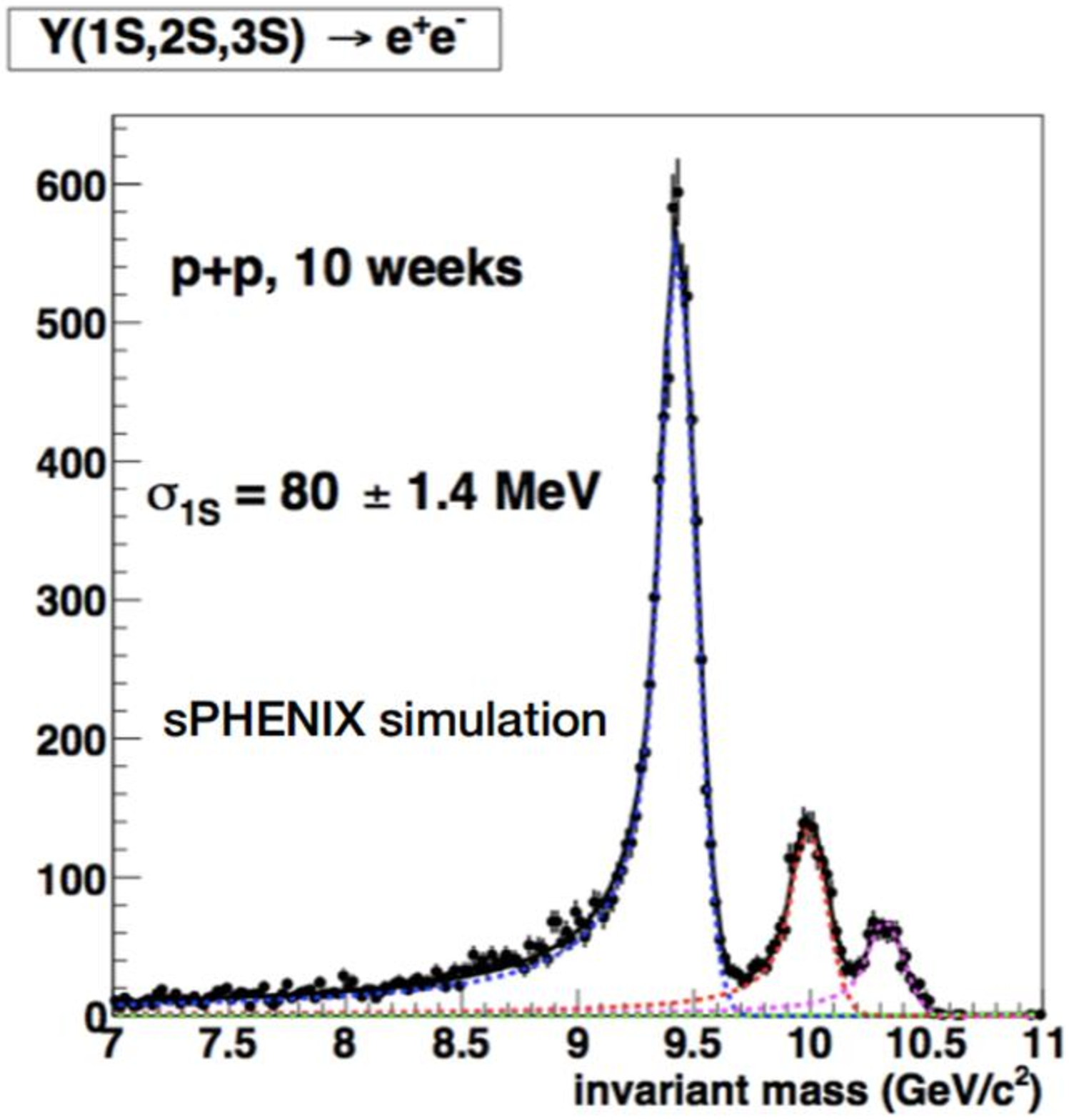}
\caption{\label{Fig:Upsilon}Simulated results of sPHENIX's ability to measure the three upsilon states from the dielectron decay channels in $10$~weeks of $p$+$p$ collisions.}
\end{minipage}
\end{center}
\end{figure}

To estimate the available jet yields in sPHENIX, projected RHIC luminosities and perturbative QCD (pQCD) rates of hard processes are needed.  Thanks to RHIC luminosity upgrades, over one hundred billion minimum bias events are expected in $22$~weeks of $\sqrt{s_{NN}}=200 GeV$ $Au$+$Au$ collisions, of which  approximately twenty billion are from $0$-$20\%$ central events.  These values combined with pQCD rate calculations provide estimates of the expected jet yields available to sPHENIX as presented in Table~\ref{Tab:pQCDyields}, confirming that rare jet probes can be measured with high statistics at sPHENIX.  Figure~\ref{Fig:RAA} shows the projected statistical errors and kinematic reach of the nuclear modification factor, $R_{AA}$, for jets, b-jets, and direct photons, assuming $22$~weeks of $Au$+$Au$ collisions and $10$~weeks of $p$+$p$ collsions at RHIC.  These projections are shown as an extension of the current RHIC capabilities with the PHENIX experiment.  Specifically, inclusive jet measurements will extend out to $80$~GeV/c in $p_{T}$, b-jet measurements will extend out to over $40$~GeV/c and direct photon measurements will extend out to over $50$~GeV/c.  With this increased kinematic range there will be significant overlap in the accessible $p_{T}$ range for jet and heavy flavor measurements at sPHENIX, the future RHIC experiment, and at future upgraded LHC experiments.
\begin{table}[h]
\caption{\label{Tab:pQCDyields}Projected yields for $0$-$20\%$ $\sqrt{s_{NN}}=200$~GeV $Au$+$Au$ collisions in $22$~weeks at RHIC.}
\begin{center}
\begin{tabular}{llll}
\br
Signal & $p_{T}$ range & pQCD & Yields\\
\mr
Light q + g jets & $p_{T} > 20 GeV/c$ & NLO & $10^{7}$\\
Light q + g jets & $p_{T} > 30 GeV/c$ & NLO & $10^{6}$\\
Direct photons & $p_{T} > 20 GeV/c$ & NLO & $10^{4}$\\
c-jets & $p_{T} > 20 GeV/c$ & FONLL & $10^{4}$\\
b-jets & $p_{T} > 20 GeV/c$ & FONLL & $10^{4}$\\
\br
\end{tabular}
\end{center}
\end{table}

To complete the b-jet program, sPHENIX needs to be able to identify or tag b-jet events.  Three methods of b-jet tagging are being explored: a) identifying multiple tracks with a large DCA, b) finding a secondary vertex, and c) tagging B-mesons by semi-leptonic decay or the baryon mass.  While method c) is still under development, Pythia8 simulations have shown that methods a) and b) can identify b-jets with an estimated $30\%$ purity and $70\%$ efficiency.  With this level of b-jet tagging, sPHENIX will be able to measure the $R_{AA}$ of b-jets out to $40$~GeV/$c$ in $p_{T}$ and potentially constrain b-jet transport coefficients in models.

For the $\Upsilon$ program, the projected upsilon yields in $10$~weeks of $p$+$p$ collisions measured in sPHENIX are shown in Figure~\ref{Fig:Upsilon}.  Over $8800$ $\Upsilon(1S)$, $2200$ $\Upsilon(2S)$, and 1160 $\Upsilon(3S)$ are expected.  Furthermore, sPHENIX can clearly separate the three upsilon states with a $\Upsilon(1S)$ width of $80\pm1.4$~MeV/$c^2$.  sPHENIX will provide, for the first time, the ability to separate all three upsilon states at RHIC.  This mass resolution is maintained in $Au$+$Au$ collisions, allowing for the centrality dependent measurement of the $R_{AA}$ in each of the upsilon states.

\section{Conclusions}

The sPHENIX project will extend RHIC results beyond PHENIX and STAR's current capablilites, and provide necessary complementary measurements to the LHC experiments.  This complementarity is needed to form a complete picture of the properties and behavior of the QGP created in heavy ion collisions.  sPHENIX will achieve this by focusing on jet, upsilon and b-jet observables.  Having recently obtained CD-0 designation, with an expected installation in 2021 and first beam available in 2022~\cite{DoE}, the future for heavy ion physics with sPHENIX is bright.

\section*{References}
\bibliography{sPHENIX_HotQuarks2016}

\end{document}